\documentclass{article}

\def \ni{\noindent}

\def \mn{\medskip
\ni}
\newcommand{\Ref}[1]{(\ref{#1})}
\def \tr{\textrm}
\def \subs{\subsection}

\def \f{\frac}

\def \tl{\tilde}

\def \beq{\begin{equation}}
\def \eeq{\end{equation}}
\def \beqs{\begin{eqnarray}}
\def \eeqs{\end{eqnarray}}

\def \harr{\hookrightarrow}
\def \I{{\cal I}}
\newcommand{\su}[1]{| #1 \rangle}
\newcommand{\Rb}{{\rm \bf R}} 
\newcommand{\Cb}{{\rm \bf C}} 
\newcommand{\Nat}{{\rm \bf N}} 

\begin{document}
    
\title{Projected Spin Networks for Lorentz connection: \\
Linking Spin Foams and Loop Gravity}
\author{{\bf Etera R Livine} \\
Centre de Physique Th{\'e}orique, \\
Luminy, Case 907, Marseille, France \\
livine@cpt.univ-mrs.fr}
\maketitle

\begin{abstract}

In the search for a covariant formulation for Loop Quantum Gravity,
spin foams have arised as the corresponding discrete space-time
structure and, among the different models,
the Barrett-Crane model seems the most promising.
Here, we study its boundary states and introduce cylindrical functions
on both the Lorentz connection and the time normal to the studied
hypersurface. We call them projected cylindrical functions and we
explain how they would naturally arise in a covariant
formulation of Loop Quantum Gravity.

\end{abstract}



\section{Introduction}

Constructing an explicitely covariant formulation of Loop
Quantum Gravity (see \cite{thiemann} for a review)
seems a natural step towards solving
some issues such as the physical role of the Immirzi parameter,
which is likely to disappear in a fully covariant approach
\cite{sergei1},
or the exact form of the hamiltonian constraint. When
studying the observables of the theory, the main difficulty
is to deal with a non-compact gauge group, the Lorentz group.
Although spin networks states have been defined for non-compact
groups \cite{spinnet}, much still needs to be done before
getting a full consistent theory. In the present paper, another
type of spin networks is defined which can be used for
non-compact gauge groups but which only have a compact
gauge symmetry. To understand where they come from, we
should review the different approaches to understanding
the physical content of the (generalised) Palatini action
\cite{holst}~:

\beq
S=\f{1}{2} \int \epsilon_{\alpha\beta\gamma\delta}
e^\alpha \wedge e^\beta \wedge (F^{\gamma\delta}+
\frac{1}{\gamma}\star F^{\gamma\delta})
\eeq
where $\alpha,\dots$ are internal Lorentz indices,
$e$ is the tetrad field, $F$ is the curvature
of the space-time connection, $\star$ is the Hodge
operator
$\star F^{\alpha\beta}=
\f{1}{2}\epsilon^{\alpha\beta}{}_{\gamma\delta}
F^{\gamma\delta}$ and $\gamma$ is the Immirzi parameter.
The main problem occuring in the canonical analysis
of this action is the second class constraints.
There have been four main paths explored in order
to deal with them.

\begin{itemize}
\item {\it Solving the constraints: Loop Quantum Gravity} \\
Holst showed it is possible to derive the canonical structure
of loop gravity with Immirzi parameter by considering a generalised
Palatini action which is related to the original by a canonical
transformation \cite{holst}. Barros pushed this analysis further by
showing that, starting off with this new action, it is possible
to solve the arising second class constraint \cite{barros}.
Then the variables
split into two couples $(A,E)$ and $(\chi,\zeta)$. The first couple
of canonical variables is a generalisation of the Ashtekar-Barbero
connection and the triad. The new variable $\chi$ is the time
normal or internal time direction and is given as the normalised
space component $\alpha=a$ 
(in the internal Lorentz indices) of the time
component of the tetrad field:

\beq
\chi^a=-\f{e^{a0}}{e^{00}}
\eeq
and $\zeta$ is its associated canonical momentum.
Finally, it is possible to gauge fix $\chi= 0$ using (the boost)
part of the Lorentz gauge symmetry.
This is called the {\bf time gauge}. And, in that frame,
we retrieve exactly the variables and constraints of loop gravity.
However, the price of this result is the loss of explicit covariance
of the theory.

\item {\it Complex Loop Quantum Gravity:} \\
In this approach by Ashtekar, which corresponds to
taking $i$ as choice for the Immirzi parameter,
we work with a complex connection and complex triad
(for a review, see for example \cite{carlo}).
In order to retrieve real results, we must impose {\bf reality
conditions}, which are indeed
the second class constraints of the theory. These conditions
constrain the measure of the Hilbert space in the quantum
theory.

\item {\it Covariant Loop Gravity:} \\
In this approach \cite{sergei1,sergei2,sergei3,sergei4},
one derives the Dirac bracket
to take into account the second class constraints directly in the
symplectic structure. This allows to still work with a
Lorentz connection. The resulting symplectic structure
ignores the space part of the connection/triad
and one gets an area spectrum depending only on the boost
part of the connection \cite{sergei2}.
This result hints toward a $SO(3,1)/SO(3)$ coset structure
in the construction of the observables.
The split space/boost
is done according to the $\chi$ field which becomes a configuration
variable that one must take in account when constructing the states
of the theory \cite{sergei4}. Despite of these results,
the drawback of this approach is that the connection is non-commutative
at the classical level. Nevertheless, on one hand,
it seems feasible to construct
commutative states for the theory \cite{sergei4}, and on the other hand,
it is possible to find a commutative Lorentz connection, which turns
out to be a covariant generalisation of the Ashtekar-Barbero connection
\cite{su2fromcov}.

\item {\it Spin Foams and the Barrett-Crane model:} \\
The spin foam approach aims toward constructing a space-time
model, which would correspond to the space analysis carried out
in Loop Quantum Gravity. This would clarify the role of time and
the form of the Hamiltonian constraint. The most developped model
is up to now the Barrett-Crane model \cite{bc1,bc2} (see
\cite{daniele} for a review).
It relies
on the (geometric) quantization of a bivector field \cite{bc1,bb,etera2}.
From the point of view of the second class constraints,
it corresponds to considering them directly at the quantum level
\cite{etera1,su2fromcov}.
They are then implemented as simplicity constraints on
the configuration space of the quantum bivector field.
These simplicity constraints amount to restrict the representations
of the Lorentz group $SO(3,1)$ in which the bivector can live. They must
have $SO(3)$ invariant subspace or, in other words, they must have
a realisation in the space $L^2(SO(3,1)/SO(3))$ \cite{bc2}.
Such representations are called simple.
The resulting boundary states of the theory are $SO(3,1)$ spin networks
labelled by simple representations and with simple intertwiners between them
\cite{bc1,bc2,finite}.
Such functionals are called {\bf simple spin networks}.

\end{itemize}

\medskip

The goal behind the present work is to make a
link between these different approaches, more particularly between
the spin foam approach and the loop quantum gravity formalism
since we would like the spin foam model to help us derive the
{\bf right dynamics } for spin networks.
Nevertheless, one is formulated using a Lorentz connection and
the other a $SU(2)$ connection.
Therefore, one issue would be to understand how get one
from the other or, in other words, to understand how
the gauge fixing procedure from $SO(3,1)$ to $SO(3)$ works at 
the quantum level. In the present work,
we first introduce {\bf projected} cylindrical functions of
the connection and the $\chi$ field, which depend
only on a finite number of arguments taken from those
fields, and which can be seen a covariant generalisation
of $SU(2)$ spin networks.
We show that the boundary states of the Barrett-Crane model,
the simple spin networks, can be easily understood in the new
context as a particular subset of projected cylindrical
functions. From this viewpoint,
the study of the Hilbert space of these new functionals
can be considered as a first step toward understanding the
quantum geometry of spin foams and its link with the canonical
formalism. Indeed the Barrett-Crane model is better understood using
the time normal and functionals cylindrical in
the time normal turn out to be a useful tool in explaining the dynamical and causal
structures of the spin foam space-time \cite{etera3}.

Moreover, the same projected cylindrical functions are shown to occur
in the covariant loop gravity formalism and we explain how they would help
understand this framework
and find a suitable hilbert space for its quantum theory.
Neverthless, the relation between this covariant approach
and the usual $SU(2)$ loop quantum gravity approach will be studied
more in details elsewhere \cite{su2fromcov}.

\section{Projected Cylindrical Functions}

\subs{A New Gauge Invariance and its physical meaning}

In the loop quantum gravity approach,
we work on a three dimensional space-like hypersurface
$\Sigma$ -the space- embedded
in a four dimensional space-time ${\cal M}$ and
we build states of the
connection, which will give the geometry of $\Sigma$
at the end of the day.
To this purpose, one usually works with cylindrical
functions. These are functions, defined on given graphs, which depend
on the connection through its holonomies on the edges of the graph.
Moreover, one require a gauge invariance of these functions which
corresponds to the invariance of
the connection state under gauge transformation
of the connection.
Precisely, given a particular graph with $E$ oriented edges and $V$ vertices
and a gauge group $G$, the gauge symmetry reads

\beq
\forall k_i\in G, \,
\phi(U_1,U_2,\dots,U_E)=
\phi(k_{s(1)}U_1k_{t(1)}^{-1},\dots,k_{s(E)}U_Ek_{t(E)}^{-1})
\label{oldsym}
\eeq
where $s(i)$ and $a(i)$ are respectively the source vertex
and the target vertex of the $i$ edge. 
All a machinery has been developped to study the structure of
such functions for compact groups such as $SU(2)$, which is the
gauge group of loop quantum gravity.
We can introduce a measure -the Haar measure-
on the space of such functions
(given a particular graph) and we can define the spin networks
as a basis of the resulting $L^2$ space. Then, we can glue
all the different spaces corresponding to all the graphs
into a global space of quantum connection state space, which
can be viewed as a $L^2$ space on a connection space
provided with the Ashtekar-Lewandowski measure. 
However, if we want to develop a covariant formalism, we are bound to use
the Lorentz group $SO(3,1)$, which is non-compact, as gauge group.
Using directly the Haar measure on such functions 
gives an infinite result and defining a new invariant measure is not
straightforward \cite{spinnet}.

In the present work, we follow the covariant approach as explained
in the introduction and we would like to use functionals of both
the (Lorentz) connection and the time normal $\chi$
to the studied hypersurface. Since it appeared useful to consider
cylindrical functions of the connection in the standard approach,
we propose in our framework to use cylindrical functions of both
the connection and $\chi$.
More precisely,
given a graph $\Gamma$ with $E$ edges and $V$ vertices, we
consider functions depending on the holonomies $U_i\in SO(3,1)$
of the connection  along the edges and the time normals
$x_i\in SO(3,1)/SO(3)$ at every node of the graph. Note
that $SO(3,1)/SO(3)$ is isomorphic to the upper sheet
of the two-sheet hyperboloid in Minkovski space.
Considering their transformation under a gauge transformation, a gauge
invariant state will be a function with the symmetry:

\beq
\label{newsym}
\forall k_i\in SO(3,1), \,
\phi (U_1,U_2,\dots,U_E,x_1,\dots,x_V)=
\phi(k_{s(1)}U_1k_{t(1)}^{-1},\dots,k_{s(E)}U_Ek_{t(E)}^{-1},
k_1.x_1,\dots, k_V.x_V)
\eeq
Due to the newly introduced $x$ variables, even though the functional
$\phi$ is invariant under $SO(3,1)^V$, the effective gauge symmetry
of $\phi$ is compact.
More precisely, it is $\bigotimes_{i=1}^{V}SO(3)_{x_i}$.
From a physical point of view, the embedding of the hypersurface
$\Sigma$ in ${\cal M}$ is defined by the whole field $\chi$. But
we have forgotten all this information to retain only the value
$x_1,\dots, x_V$ of $\chi$ at the vertices of the graph so that
the embedding of $\Sigma$ is defined at only those points while
it is left fuzzy everywhere else. And imposing the normal at a vertex
to be $x_i$ implies that
the symmetry at this vertex is reduced
from $SO(3,1)$ to $SO(3)_{x_i}$. And it is this effective gauge symmetry
that will have to be taken into account when constructing the measure, as we
will see in the next paragraph.

\medskip
A natural choice for the $x$ variables is the vector $x_0=(1,0,0,0)$.
This corresponds to the time gauge $\chi=0$ and it is the usual
gauge fixing that one does when carrying the canonical analysis
of the Palatini action through a $3+1$ splitting. In our framework,
we don't fix the $x$ variables to a given value but we allow them to vary.
This would allow an analysis without gauge fixing and thus explicitely
covariant. Moreover, it opens a door to analysing changes of gauge
fixing such as the study of a Lorentz boost (see \cite{simone}
in the case of Loop Gravity) through the shift
$x_1=\dots=x_V=x_0$ to $x_1=\dots=x_V=(1,v,0,0)$.

\subs{The $L^2$ space of Projected Cylindrical Functions}

We would like to give a Hilbert space structure to
the space of cylindrical functions defined previously
in order to raise them to the status of quantum
states of the hypersurface. Indeed, we can use the simple
Haar measure $dg^E$ to integrate functions with the gauge symmetry
\Ref{newsym}. And we define the Hilbert space of
Projected Cylindrical Functions as the space of $L^2$ functions
satisfying \Ref{newsym}. This induces the scalar product:

\beq
\langle \phi | \psi \rangle=
\int_{SO(3,1)^E} \prod_i\tr{d}g_i
\bar{\phi}(g_1,\dots,g_E,x_1,x_2,\dots,x_V)
\psi(g_1,\dots,g_E,x_1,x_2,\dots,x_V)
\label{newprod}
\eeq

\ni
One can check that this definition doesn't depend
on the choice of $x_1,\dots,x_V$.
Nevertheless, we have to consider both functions
$\phi$ and $\psi$ with the same $x$ arguments i.e to lay
in the same embedded hypersurface.
Still, this leaves us a freedom in the choice of the $x$'s and
we could fix them all to the unit time vector $x_0=(1,0,0,0)$ 
which is the time gauge.

\medskip
It is possible to simply use the Haar measure $dg^E$ because
the effective gauge invariance of functions satisfying \Ref{newsym}
is compact, being $SO(3)^V$ in the time gauge. What we have done is
cancel the non-compact part $SO(3,1)/SO(3)$ of the gauge symmetry.
More precisely, it is possible to re-construct the whole function $\phi$
from the function

\beq
\tl{\phi}(g_1,\dots,g_E)=\phi(g_1,\dots,g_E,x_0,x_0,\dots,x_0)
\eeq
which satisfies a $SO(3)$ gauge invariance:
\beq
\forall k_i\in SO(3)_{x_0}, \,
\tl{\phi}(g_1,g_2,\dots,g_E)=
\tl{\phi}(k_{s(1)}g_1k_{t(1)}^{-1},\dots,k_{s(E)}g_Ek_{t(E)}^{-1})
\eeq

\ni
We can visualise this space of functions on a scale of structures:

\begin{itemize}
\item There is the space of non-gauge invariant functions, which
don't satisfy any symmetry requirement. We can construct a Hilbert space
of such functions by simply considering the Haar measure and we get
${\cal H}_{\mathrm non-inv}=L^2(dg^E)$.
\item There is the presently introduced $SO(3)$ invariant functions which
can be obtained from the previous set of functions by integrating
over the $SO(3)$ part of the group elements. We can still use the Haar measure
and construct the Hilbert space
${\cal H}_{{\mathrm proj}}=L^2_{SO(3){\mathrm inv}}(dg^E)$.
\item There is the full gauge invariant functions which can be obtained
from the previous set by integrating over the remaining $SO(3,1)/SO(3)$
symmetry or equivalently over the $x$ variable. However, we can not use
directly the Haar measure and we must construct a quotient measure which
make harder to deal with the corresponding Hilbert space \cite{spinnet}.
\end{itemize}

\ni
We should note that we can still implement gauge-invariant observables
on the space ${\cal H}_{{\mathrm proj}}$ in a finite and gauge-invariant way.
Indeed, let's consider a gauge invariant function
$A(g_1,\dots,g_E)$ invariant under \Ref{oldsym}.
One can consider the matrix elements of $A$~:

\beq
\langle\phi|A|\psi\rangle=
\int_{SO(3,1)^E} \prod_i\tr{d}g_i
\bar{\phi}(g_1,\dots,g_E,x_1,\dots,x_V)
A(g_1,\dots,g_E)
\psi(g_1,\dots,g_E,x_1,\dots,x_V)
\label{scalar}
\eeq
This doesn't depend on the choice of the $x$'s and is finite as
long as $A$ is bounded.
The main issue is then to physically interpret the role of the $x$ variables.

\subs{Projected Spin Networks}

It is possible to find an orthonormal basis to the introduced $L^2$ space.
They will have a similar structure as spin network but the symmetry of the
intertwiner at a vertex $i$ will be projected
from $SO(3,1)$ down to $SO(3)_{x_i}$,
so that we call the elements of this basis projected spin networks.
More precisely,
we are going to construct some
spin network which will be defined on $SO(3,1)$ but will have
$SO(3)$ intertwiners.

\medskip

Let's choose an oriented graph $\Gamma$ 
with $E$ edges and $V$ vertices embedded in the space $\Sigma$.
Let's construct the holonomies of the connection $A$ along the edges
$e_1,e_2,\dots , e_E$ and denote them $U_1,\dots, U_E$.
The normal procedure to construct a spin network would be to assign
a representation ${\cal I}_i$ of $SL(2,C)$ to each edge $e_i$, then
to choose $SL(2,C)$ intertwiners $I_v$ for every vertex $v$ of the graph and
build the spin network functional by contracting the holonomies $U_i$
in the representations ${\cal I}_i$ with the intertwiners to get a scalar:

\beq
\phi (A)= \bigotimes_v I_v \bigotimes_i D^{{\cal I}_i} (U_i)
\eeq

\ni
In our case, we further choose a $SU(2)$ representation $j_i^{(v)}$
for each edge $e_i$ of the graph - for each couple of vertex and
incident edge -
and a $SU(2)$ intertwiner $i_v$ between these representations
for each vertex $v$.
To glue the holonomies using these intertwiners, we project them at
the vertices. To be more precise, let's introduce the representation
space $R^{\cal I}$ of the representation ${\cal I}$ of $SL(2,C)$
and the representation space
$V^j$ of the representation $j$ of $SU(2)$.
Unitary irreducible representations of $SL(2,C)$ are infinite
dimensional
and labelled by a positive integer $n$ and a positive real number
$\rho$ (see appendix \ref{B} for more details).
Then, we can decompose $R^{(n,\rho)}$ into $SU(2)$
representations. For this purpose, we choose a particular $SU(2)$
subgroup of $SL(2,C)$ i.e we choose a time normal
$x \in SL(2,C)/SU(2)$ and consider the subgroup $SU(2)_x$ of
elements leaving the vector $x$ invariant (see appendix \ref{A} for more
details) . Then:

\beq
R^{(n,\rho)}=\bigoplus_{j\ge n} V^j_{(x)}
\eeq

\ni
We call $P^{(x)}_j$ the orthogonal projector from
$R^{(n,\rho)}$ to $V^j_{(x)}$. More explicitely, we
can express it as

\beq
P^{(x)}_j=\Delta_j\int_{SU(2)_x}\tr{d}g\overline{\chi}^j(g)D^{(n,\rho)}(g)
\eeq

\ni
where $\Delta_j=(2j+1)$ is the dimension of the $j$ representation,
the integration is over the subgroup $SU(2)_x$,
$D^{(n,\rho)}(g)$ is the matrix of $g$ in the $(n,\rho)$ representation
and $\chi^j$ is the
character of the $j$ representation.

\ni
To construct the projected
spin network, we insert this projector at both ends of every edges
and obtain the following structure around a (3-valent as an example)
given vertex $v$:

\beq
\phi(U_1,U_2,U_3,\dots, x_v,\dots)=
i_v^{j_1j_2j_3}\prod_{i=1}^3|{\cal I}_ix_vj_i^{(v)}m_i\rangle
\langle{\cal I}_ix_vj_i^{(v)}m_i|D^{{\cal I}_i}(U_i) \dots\tr{other vertices}
\eeq

\ni
where $|{\cal I} xjm\rangle$ is the basis of $V^j_{(x)}\harr R^{\cal I}$, $m$
running from $-j$ to $j$, and where we sum over the $m_i$.

It is easy to check that these constructed spin networks
satisfy the gauge invariance \Ref{newsym}. And a straightforward
calculations shows that, once we have chosen an orthonormal
set of $SU(2)$ intertwiners, the resulting projected spin networks
form an orthonormal basis of the $L^2$ space of projected
cylindrical functions i.e the scalar product of two spin networks
is a delta function on the representations ${\cal I}$ and $j$ and
on the intertwiners.

Let's emphasize that there is two a priori different $SU(2)$
representations on each edge, at both its ends. When we will
tackle the issue of refining the graph, we will then restrict ourself
to projected spin networks with a single $SU(2)$ representation on
every edge (see section \ref{refine}).


Let's give some examples. The simplest is the one-loop
graph with a single vertex.
We are looking for functions
$\phi$ on $SL(2,C)\otimes SL(2,C)/SU(2)$ satisfying

\beq
\forall k\in SL(2,C),\,
\phi(g,x)=\phi(kgk^{-1},k.x)
\label{1loop}
\eeq
or after having gauge fixed $x$ to, let's say, the time gauge, we can write
the gauge invariance on the gauge fixed function
$\tl{\phi}(g)=\phi(g,x_0)$:

\beq
\forall h\in SU(2),\,
\tl{\phi}(g)=\tl{\phi}(hgh^{-1})
\eeq
The projected spin network construction gives functions labelled by
one $SL(2,C)$ representation ${\cal I}=(n,\rho)$ and
one $SU(2)$ representation $j$. Instead of
considering the character of the representation $(n,\rho)$, we restrict ourself
to the trace over the subspace $V^j_x$:

\beq
\phi^{(n,\rho)}_j(g,x)= \textrm{Tr}_{V^j_x}^{(n,\rho)}(g)
\eeq
Another  example is the $\Theta$ graph. It has two 3-valent vertices
$A$ and $B$. We choose three $SL(2,C)$ representations and three $SU(2)$
representations. The resulting projected spin networks is given by
taking the matrix elements of the three $SL(2,C)$ group elements in
the $SL(2,C)$ representation restricted to the spaces $V^j_x$ and contract
them with the two Clebsh-Gordon coefficients which are the unique
three-valent $SU(2)$-intertwiner (up to normalisation).

\subs{Going down to $SU(2)$ spin networks}

One interesting issue is
how these newly defined projected spin networks could
relate to the usual $SU(2)$ spin networks which are a basis of the
states of quantum geometry in the usual formulation of loop quantum
gravity.

In fact, the projected spin networks easily go down to $SU(2)$ spin 
network. More precisely, in the special
gauge where all $x$ variables are the same, let's say equal to $x_0$,
it is possible to restrict the group variables to live in the $SU(2)_{x_0}$
subgroup and consider the (projected) functionals as functions
over $SU(2)^E$. Due to the gauge invariance \Ref{newsym}, they are
effectively cylindrical functions on $SU(2)$.
Even better, projected spin networks with edges labelled
by representations $({\cal I}_e=(n_e,\rho_e),j_e)$
(the two $SU(2)$ representations at the two ends of one given edge must match, else
the resulting $SU(2)$ spin network is 0, so we choose only one $SU(2)$
representation per edge)
simply reduces to a $SU(2)$ spin network
with edges labelled by the $SU(2)$ representations $j_e$,
the vertices still labelled by the same $SU(2)$ intertwiners.
Moreover, this procedure is obviously independent from the choice of gauge as long as
all the $x$'s are chosen to be equal:
one gets the exact same $SU(2)$ spin
network.

Interestingly,
the $SL(2,C)$ representations $(n_e,\rho_e)$ coming into the definition of the
initial projected spin network don't intervene in the definition of the resulting
$SU(2)$ spin network.
Choosing other representations $(\tl{n}_e,\tl{\rho}_e)$ still gives
the same $SU(2)$ spin network (as long as we have chosenthe same $j_e$).
Thus, they don't seem to describe the 3d geometry
but more how it changes under Lorentz boost:
they can be understood as the space-time embedding
of the studied space slice (or the structure of the space-time around that slice).
This is a reason why the present formalism leads to a covariant definition
of the $SU(2)$ spin networks.

\subs{Theory for arbitrary group}

We start by recalling some group theory to introduce the maximal compact
subgroup of a group $G$ and its basic properties. The 
following elements are taken from \cite{knapp}.
We consider a linear connected reductive group $G$
i.e a closed connected group of real or complex matrices that is
stable under conjugate transpose. We note ${\cal G}$ its Lie algebra.
The inverse conjugate transpose, denoted $\Theta$, is an automorphism of $G$
and is called the {\it Cartan involution}. We note that $\Theta^2=1$.
Let

\beq
K=\{g\in G | \Theta g=g\}.
\eeq

\ni
$K$ is a compact connected  and is a maximal compact subgroup of $G$.
For example, it is $SO(3)$ in the case of $G=SO(3,1)$ and\
$SU(2)$ in the case of $G=SL(2,\Cb)$.
The differential $\theta$ of $\Theta$ at the identity $1$ is an
automorphism of ${\cal G}$, given as the negative conjugate transpose. 
Since $\theta^2=1$, we have the {\it Cartan decomposition} for ${\cal G}$

\beq
{\cal G}=p\oplus m\, ,
\eeq

\ni
where $p$ and $m$ are the eigenspaces corresponding to
the eigenvalues 1 and -1 of $\theta$. $p$ consists of the
skew-Hermitian members of ${\cal G}$, $m$ of the Hermitian elements,
and it is easy to derive the following relations:

\beq
\begin{array}{ccc}
\left[p,p\right] & \subset & p \\ 
\left[p,m\right] & \subset & m \\ 
\left[m,m\right] & \subset & p
\end{array}
\eeq

\ni
The {\it Cartan decomposition} for $G$ is the map

\beq
\begin{array}{ccc}
K\times m &\rightarrow & G \\
(k,u) & \rightarrow & k \,\exp(u)
\end{array}
\eeq

\ni
It is a diffeomorphism onto $G$, which identifies the quotient
$X=G/K$ of the group $G$ under the left action of $K$ to the space
$m$.

\mn
A {\it projected cylindrical function} will be defined on a
graph $\Gamma$ - with $E$ oriented edges and $V$ vertices -
and will depend on a group element for each edge
and a spin variable $x\in G/K$ for each vertex and will have
the gauge symmetry:

\beq
\forall k_i\in G, \,
\phi (g_1,g_2,\dots,g_E,x_1,\dots,x_V)=
\phi(k_{s(1)}g_1k_{t(1)}^{-1},\dots,k_{s(E)}g_Ek_{t(E)}^{-1},
k_1.x_1,\dots, k_V.x_V)
\eeq
Such a function can be constructed from an arbitrary
function over $G^{\otimes E}$ by (left-)integrating over the $K$
part of the $E$ group elements.
The gauge fixing corresponds to fixing all the $x$ variables to
some arbitrary
values and the {\it time gauge} amounts to
fix all $x$ to the equivalence class $[\textrm{Id}]$ of the identity
i.e the group $K$ itself. 
The measure we introduce is the Haar measure $dg^{\otimes E}$.
The resulting Hilbert space is the space of $L^2$ functions satisfying
the required symmetry.
And we see that we have managed to sidestep the problem of dealing with a
non-compact gauge group.

\medskip

We now would like to also generalise the construction of the
projected spin networks. To this purpose, let's see
how a representation of $G$ decomposes over representations of $K$.
To be more precise,
let $G$ be a linear connected reductive group and $\pi$ a representation
of $G$ on a Hilbert space $V$. When $K$ acts by unitary operators, then
we can decompose $\pi$ into irreducible representations of $K$
\cite{knapp} (chapter 8):

\beq
\pi \, = \, \sum_{\tau\in\hat{K}} n_\tau \tau
\eeq

\ni
where $\hat{K}$ is the set of equivalence class of
irreducible representations of $K$ and
$n_\tau\in\Nat\cup\{+\infty\}$ is the
multiplicity of each representation.
The equivalence classes $\tau$ that have a positive
multiplicity are called {\it $K$ types}.
Moreover, if $\pi$ is an irreducible unitary representation,
all the multiplicities $n_\tau$ are finite and satisfy
$n_\tau \le \textrm{dim}\tau$.

\ni
We can further introduce $K$-finite vectors which
are vectors $v$ such that $\pi(K)v$ spans a
finite-dimensional space. Then, for unitary
representations (or more generally
so-called admissible representations),
all $K$-finite vectors are $C^\infty$ vectors, the
space of $K$-finite vectors is stable under
$\pi({\cal G})$ and every matrix coefficient
$g\rightarrow (\pi(g)u,v)$, with $u$ $K$-finite,
is a real analytic function on $G$.

\ni
We are then ready to introduce the projected spin networks.
We choose $E$ group elements $g_1,\dots, g_E$ on the edges
of the graph, and $V$ elements $x_1,\dots,x_V$ of the coset
$G/K$. We further choose $E$ representations ${\cal I}$
of $G$, and 2$E$ representations of $K$. We can follow
the exact same construction as in the $SL(2,C)$ case,
choosing $V$ $K$-intertwiners and contracting them using
the matrix element of $\pi^{{\cal I}_i}(g_i)$ in
a basis of the representation space of the $K$-subgroups
corresponding to the choice of $x$'s. Finally, due to the
$K$-finiteness of the vectors used in the construction,
the resulting function is perfectly well-defined.

\section{The Barrett-Crane model and its boundary states}

\subs{Simple Spin Networks}

The Barrett-Crane model is a spin foam model. More precisely,
it is a theory of space-time based on a simplicial
discretisation with representations of $SL(2,C)$ living
on the faces (triangles) of the decomposition (see \cite{daniele}
for a review). Its boundary states are 4-valent
simple spin networks
i.e $SL(2,C)$ spin networks with simple representations
and simple intertwiners. The link of the model with general
relativity is that it appears as a discretisation of
a constrained BF theory equivalent to the usual
Palatini action \cite{freidel,etera2}.

There exists two Lorentzian models. The first one \cite{alejl1}
is based on simple representation that can be represented
on the space of $L^2$ functions over the 2-sheet hyperboloid
$SL(2,C)/SU(2)$.
The induced representations are of the type $(0,\rho)$
(see appendix B for a review of the irreducible unitary
representations of $SL(2,C)$). The second one
\cite{alejl2} uses the space of $L^2$ functions over
the one-sheet hyperboloid $SL(2,C)/SU(1,1)$
and induces all the simple
representations $(n,0)$ and $(0,\rho)$. They can both
be expressed in term of a field theory over $SL(2,C)$.
In the present work, we will deal only the first model
because the structure of its boundary term allows a
simple interpretation in terms of space-like hypersurface,
as we will see later.

We are now going to have a look at the simple spin networks
that arise in this Lorentzian model and explain how
they enter in the framework of projected spin networks.
Here, we will only review the elements necessary to the
present analysis. \cite{simple} gives a general presentation of simple
spin networks and one can look in \cite{bc2,finite} for
a precise account of Lorentzian simple spin networks.
The results are as follows.
Simple spin networks are defined by simple representations
$\rho_e$
($SL(2,C)$ irreducible representations
containing a $SU(2)$ invariant subspace)
living on the edges of the graph. At the vertices,
we contract the matrix elements of the group elements
using simple intertwiners. The interesting thing
is that these intertwiners can be written as an
intregral over $SL(2,C)/SU(2)$. More precisely,
to compute the simple spin network, one chooses
a $x_v\in SL(2,C)/SU(2)$ for each vertex.
Then, for each edge $e$, one computes the corresponding
Kernel (as defined in the appendix \ref{C})

\beq
K_{\rho_e}(x_{s(e)},g_ex_{t(e)})=
\langle
\rho_e (j=m=0)|(x_{s(e)})^{-1}g_ex_{t(e)}|\rho_e (j=m=0)
\rangle
\eeq
Finally one considers the product of these Kernels and
integrate over the $x$ variables. It appears that if one forgets
about the final integration, one has exactly computed
the projected spin networks defined earlier in the special
case when one sets the $SU(2)$ representation to the
trivial one $j=0$ everywhere. In particular, one can interpret
the $x$ variables as the time direction and therefore the
simple spin networks as describing a space-like hypersurface.
If one had considered the simple spin networks for $SL(2,C)/SU(1,1)$,
one would have been correspondingly describing a time-like
hypersurface.

As projected spin networks, the simple spin networks
satisfy the symmetry \Ref{newsym}. Nevertheless, they have an
extra symmetry: at each vertex, they are invariant under $SU(2)^n$
instead of a simple $SU(2)$, $n$ being the number of edges
linked to that given vertex. In the spin foam framework,
this comes from the simplicity constraints i.e corresponds
to the implementation of the second-class constraints.

One can easily check the orthonormality of the simple spin networks
for the scalar product \Ref{scalar}
using the formula of appendix \ref{C}. Indeed, when computing the
scalar product of two simple spin networks, once we have fixed
the $x$ variables at all the vertices, we can split the integral
over $SL(2,C)^E$ into the product of $E$ integrals over $SL(2,C)$
of the product of the two Kernels corresponding to the same edge
but to the two different spin networks, then

\beq
\int_{SL(2,C)}dg
K_\rho(x_{d(e)},gx_{t(e)})K_{\rho'}(x_{d(e)},gx_{t(e)})=
\f{\delta(\rho-\rho')}{\rho^2}K_\rho(x_{d(e)},x_{d(e)})=
\f{\delta(\rho-\rho')}{\rho^2}
\eeq
which shows the orthonormality of the simple spin networks.
A simple cylindrical function is obtained by integrating
over the representations $\rho$
the defined simple spin network with $L^2(\rho^2d\rho)$
distributions using the Plancherel measure $\rho^2d\rho$.
As an example, let's consider a single edge and the corresponding
simple spin network $K_\rho(x,gy)$. A $L^2$ function $f$ is defined
by a $\alpha\in L^2(\rho^2d\rho)$ as

\beq
f(g,x,y)=
\int \rho^2d\rho \, \alpha(\rho) K_\rho(x,gy)
\eeq
and its $L^2$ norm is the $L^2$ norm of the function $\alpha$.

Let's also point out that the function given by a
single Kernel $f(g,x,y)=K(x,gy)$ is a projected cylindrical
function: it is possible to consider {\bf open}
spin networks (which are still gauge invariant)
within the context of projected cylindrical spin networks.

\subs{3d geometry from the spin foam boundary?}

An interesting question in order to understand
the physical content of these simple spin networks
is what kind of ``quantum'' 3d geometry do they describe.
Indeed, they are supposed to be the quantum state
of the hypersurface. So what data do they contain?

First, one can look at the $SU(2)$ spin networks induced by
the simple spin networks following the same procedure
as in the previous section: let's choose a gauge where
all the $x$'s are fixed to $x_0$ and study the functions restricted
to $SU(2)_{x_0}$ group elements.
Then, one finds $SU(2)$ spin networks
with edges labelled by the trivial representation $j=0$: the
simple spin networks don't seem to carry any information
on the $SU(2)$ restriction of the connection.
Indeed,  the simple spin networks
can be considered as the Fourier
decomposition of the requirement that the $SL(2,C)$
group element live in $SU(2)$. 
More precisely, one has the equation:

\beq
\delta_{{\cal H}_+}(x_{d(e)},g_ex_{a(e)})=
\f{1}{2\pi^2}
\int_0^\infty \rho^2d\rho\,K_\rho(x_{d(e)},x_{a(e)})
\eeq
which says that the $K_\rho(x_{d(e)},x_{a(e)})$ 
(one edge of a simple spin network) can be seen as the Fourier
transform of the requirement that
$g_e$ must define a parallel transport, along the edge $e$,
which takes $x_{d(e)}$ to $x_{a(e)}$.
However, there exists a whole $SU(2)$ subgroup
of such group elements among which $g_e$ can be randomly chosen. 

Due to this, simple spin networks
don't give amplitudes differentiating between
the $SU(2)$ part of the group elements:
they are not similar
to using $SU(2)$ spin networks, which are interpreted as
giving an amplitude on the $SU(2)$ connection defining the 3d
geometry. On the other hand, they describe how the space hypersurface
changes under Lorentz boosts i.e under change of embedding into
the space-time.
They seem more like an orthogonal information:
$SU(2)$ spin network seems an extra information one would
have to impose by hand.
$SU(2)$ spin networks don't seem to
arise naturally in the Barrett-Crane model and
that's why it seems more natural to compare
the spin foam setting with the covariant canonical analysis
of the Palatini action as done by Alexandrov
\cite{sergei1,sergei2,sergei3,sergei4}.

Still, thinking the Euclidean Barrett-Crane model
based on the group $Spin(4)=SU(2)_L\otimes SU(2)_R$,
simple representations (which contain a vector invariant
under the diagonal $SU(2)$ subgroup)
are representations $(j,j)$ (same representations for $SU(2)_L$
and $SU(2)_R$) and
simple spin networks labelled by $(j_e,j_e)$ are thought to
go down to $SU(2)$ spin networks labelled by the representations $j_e$
(because the areas defined by the Casimirs is the same for such a choice):
we don't restrict to the diagonal $SU(2)$ which is the space symmetry group
but to $SU(2)_L$ (or $SU(2)_R$ which would be equivalent). Therefore, maybe
a $SU(2)$ spin network interpretation of Lorentzian spin networks is still
possible considering another type of restriction of the group elements.
A more careful study of the gauge fixing procedure would be needed for such
purpose.

\section{Towards Covariant Loop Quantum Gravity?}
\label{covariant}

\subs{Fock Space of Projected Cylindrical Functions}

At the end of the day,
one would like to have a Hilbert space of
quantum states of geometry. And more
precisely, in loop-like approaches, we
are looking for a Hilbert space of quantum
states of the connection, which should be obtained
by summing over all the spin networks states i.e
summing the Hilbert spaces corresponding to all the
different graph. In Loop Quantum Gravity,
through the Ashtekar-Lewandowski construction,
it is possible to make this space appear
as a $L^2$ space \cite{al}. This even works
for any compact group. But this approach fails
in the case of non-compact groups \cite{spinnet}.

In the case of projected cylindrical functions,
it is nevertheless possible to give a Fock space structure
to the sum over graphs. This is due to the use of the
normal $SL(2,C)$ Haar measure (and not a quotient
measure as in \cite{spinnet}) allowed by the
compact effective gauge invariance. More precisely,
let's choose a graph $\Gamma$ and a subgraph
$\Gamma_1$, a $L^2$ cylindrical function $f$
on $\Gamma$ and $L^2$ cylindrical function
$\varphi$ on $\Gamma_1$.
The action of the annihilation operator $a_\varphi$
on the function $f$ will give a $L^2$ cylindrical
function on $\Gamma'=\Gamma\setminus\Gamma_1$ defined as
the set of edges of $\Gamma$ which are not in $\Gamma_1$
linked by the needed vertices of $\Gamma$ (vertices which are
not in the interior of $\Gamma_1$:

\beq
a_\varphi f (g_{f\in\Gamma'},x_{w\in\Gamma'})=
\int \prod_{e\in \Gamma_1}\tr{d}g_e
\bar{\varphi}(g_{e\in\Gamma_1},x_{v\in \Gamma_1})
f(g_{e\in\Gamma_1},g_{f\notin\Gamma_1},x_{v\in\Gamma})
\eeq
where the $x_{v\in \Gamma_1}$ variables are taken to be the same
for $\varphi$ and $f$.

We can also write creation operators acting on a given graph $\Gamma$,
adding edges to it by acting on $L^2_{SO(3){\mathrm inv}}(\Gamma)$
with a cylindrical function depending only on the added edges and the
corresponding vertices. This action amounts to multiply the two
functions together. One must be careful that we are {\bf not} gluing
two graphs together along some common edges, but only adding some
edges: the only thing in common are vertices. This other situation
would be more complicated and would need some convolution product,
which will not be investigated here.

Finally, we have seen it is possible to endow the sum over graphs
of all $L^2$ projected cylindrical functions of a {\bf Fock space
structure}.

\subs{Refinement and the bivalent vertex problem}
\label{refine}

In this paragraph, we will tackle the issue of refining
the projected cylindrical functions. More precisely, I mean
that given a graph, a projected cylindrical function is projected
only at the vertices of the graph and we will see how to project it at
virtually every point of the graph precising the time direction at all
points of the graph. This would correspond to (fully) projected spin
networks as introduced in \cite{sergei4} for a proposal of Hilbert
space for covariant loop gravity.

We will therefore study bivalent vertices and how to pass between
the space of functions depending on one edge and its two vertices
and the ones depending on two edges linked with a bivalent vertex
(while ignoring other edges of the graph):

\beq
\begin{array}{cccc}
f(g,x,y)&=&&f(kgh^{-1},kx,hy) \\
&  \updownarrow &? & \\
\phi(g_1,g_2,x,y,z) &= &&
\phi(ag_1c^{-1},cg_2b^{-1},ax,by,cz)
\end{array}
\eeq
The most natural operation between the two sets is to start with 
a function $f$ and define a $\phi$ function
by contracting its two group elements:

\beq
f \rightarrow \phi(g_1,g_2,x,y,z)=
f(g=g_1g_2,x,y)
\label{contract}
\eeq
One can write an ``inverse'' to this operation
to go from a $\phi$ function to a $f$ function
by integrating over $z$. More precisely,
$\int dz \,\phi(g_1,g_2,x,y,z)$
depends on $g_1$ and $g_2$ through only $g=g_1g_2$.
Moreover, we can express this same integral using a
kind of convolution:

\beq
\phi \rightarrow
f(g,x,y)=\int_{SL(2,C)} \tr{d}\tl{g}\,
\phi(\tl{g},\tl{g}^{-1}g,x,y,z)
=\int_{{\cal H}_+} \tr{d}z \,
\phi(g_1,g_2,x,y,z)
\label{inverse}
\eeq

However, this is the contrary of what we are looking for since
we would like to tell what is the value of the intermediate
variable $z$ i.e insert a dependence over $z$ instead of ignoring it as
in \Ref{contract}. We are going to see how this can be done working
on projected spin networks. To start with, let's focus on a given edge
of a projected spin network. Three variables live on this edge: a
group element $g$ and two time normals $x$ and $y$ at its two
vertices. The spin network further depends on a $SL(2,C)$
representation ${\cal I}$ and two $SU(2)$ representations
$j_1,j_2$ so that the function reads

\beq
f(g,x,y,\dots)=\langle{\cal I} x j_1 m_1|
D^{\cal I}(g)|{\cal I} y j_2 m_2\rangle \dots
\eeq
We are going to create a bivalent vertex in the middle of this edge
by inserting the Identity
\beq
\forall j,\,\tr{Id}_{R^{\cal I}}=\f{1}{\Delta_j}\int_{{\cal H}_+}
\tr{d}z P^j_{(z)}.
\eeq
This will work iff $j_1=j_2=j$. Indeed, inserting a bivalent vertex
means also imposing a $SU(2)$ invariance at this vertex. And this
imposes to have the same $SU(2)$ representation on both side.
Therefore, we are considering projected spin network with a unique
$SU(2)$ representation for each edge. Then, we can consider the
following function
\beq
\phi(g_1,g_2,x,y,z,\dots)=
\f{1}{\Delta_j}
\langle{\cal I} x j_1 m_1|D^{\cal I}(g_1)
P^j_{(z)}D^{\cal I}(g_2)|{\cal I} y j_2 m_2\rangle
\dots
\eeq
Once we use the operation \Ref{inverse} on this function,
we fall back onto our feet and find back the function $f(g,x,y)$
we started with. This way, we can precise the normal $z$ at a point
on an edge of the graph. We would like to repeat this process
infinitely many times to fully project our cylindrical function.
Nevertheless, a problem is that this ``projection'' operation is not
straightforwardly consistent with the scalar product i.e it doesn't
respect (commute with) the $L^2$ structure we are using. For the sake
of the simplicity of the notations, we will look only at simple
spin networks, but the following is also valid for any projected spin network.

Let's consider a 2-edge graph with an edge
going from $x$ to $z$ labelled with
a $L^2(\rho^2d\rho)$ representation distribution $\alpha(\rho)$
and an edge going from $z$ to $y$ labelled with $\beta(\rho)$:

\beq
\phi(g_1,g_2,x,y,z)=
\int\rho_1^2d\rho_1\int\rho_2^2d\rho_2\,
\alpha(\rho_1)\beta(\rho_2)
K_{\rho_1}(x,g_1z)
K_{\rho_2}(z,g_2y)
\eeq
$\phi$ is a $L^2$ function using the measure
defined for the 2-edge graph and its norm is

\beq
|\phi|^2 =\left(\int\rho^2d\rho \,\alpha^2(\rho)\right)
\left(\int\rho^2d\rho \,\beta^2(\rho)\right)
\eeq
We coarse grain it using \Ref{inverse} and we obtain the function $f$
defined on a 1-edge graph going from $x$ to $y$:

\beq
f(g,x,y)= \int \rho^2d\rho \,\alpha(\rho)\beta(\rho)
K_\rho(x,gy)
\eeq
The norm of $f$ using the measure defined for the 1-edge graph is

\beq
|f|^2=\int\rho^2d\rho\,\alpha^2(\rho)\beta^2(\rho)
\eeq
which is not finite in general i.e $f$ is not automatically
a $L^2$ function! Nevertheless, $f$ is $L^2$ if $\alpha$
and $\beta$ are both in $L^4(\rho^2d\rho)$. And if we want
to refine more and more, we will have to choose functions
in $L^2\cap L^4 \cap L^6 \cap \dots$, as for example
functions with compact support. It will still be possible to
refine projected cylindrical functions using the procedure
described for projected spin networks. But the $L^2$ norm
of the refined functions will not be equal.

There is another point of view, which amounts to impose
that the norm is conserved by refining. Indeed the
coarse graining projection \Ref{inverse}
is obviously generalised to graphs more and more refined
(with more and more bivalent vertices) and allow to define
projections $p_{\Gamma_1\Gamma_2}$ with $\Gamma_2\subset\Gamma_1$
($\Gamma_2$ is a coarse grained version of $\Gamma_1$: we have integrated
out some bivalent vertices) which take $L^1_{\Gamma_1}$ functions
to $L^1_{\Gamma_2}$ functions. Thus, one can define sequences of consistent
$L^1$ functions (which project on one another) whose integral
(using the Haar measure) is conserved
under refining of the graph. This means we can define a
(generalised) measure $d\mu_\infty$
over such consistent sequences by saying that the integral
of a sequence is simply the integral of one of its functions
(since these integrals are all the same).
We can also define a Hilbert space
$L^2(d\mu_\infty)$.  The drawback of this approach is that the refining
of such $L^2$ functions is not \Ref{inverse} anymore but

\beq
\phi \rightarrow f = \left(\int d\tl{g}\phi^2 \right)^{\f{1}{2}}
\eeq
which doesn't lead to the ``nice'' (easily interpreted) refining of the projected
spin network described above.

\subs{Projected Spin Networks as Area Eigenvectors}

An argument for the use of projected cylindrical functions in covariant
loop gravity is their link with Alexandrov's covariant approach to canonical
loop gravity \cite{sergei1}. More precisely, in his approach,
one deals with a $SL(2,C)$ connection $A_i^X$
($i$ being the space index and $X$ a $sl(2,c)$ index)
and a tetrad field $P_Y^j$ and
one can derive
a Dirac bracket, taking into account the second class constraint,
which projects onto the boost part of the connection:

\beq
\{ A_i^X(x),P_Y^j(y)\}_D=I_{(\chi)Y}^X\delta_i^j\delta(x,y)
\eeq
where $\chi$ is the time normal field and
the projector $I_{(\chi)Y}^X$ is 1 when $X$ and $Y$ are in the boost part
orthogonal to $SU(2)_\chi$ and 0 in all other cases
(see \cite{sergei2} for more details).

Here the projected cylindrical functions allow a discrete representation
of this commutation relation. More precisely, let's choose a finite number
of points on the manifold. Then, by considering graphs whose set of vertices
contains those points,
one gets a representation of the commutation relations at the chosen points.
Moreover, one can define the area operator at the vertices of the graph.
Indeed for each couple of vertex and incident edge, the projected spin networks
are eigenvectors of the area operator of a surface intersecting the
edge at the vertex and 
one gets the area spectrum
derived in \cite{sergei2}:

\beq
\tr{Area} \sim \sqrt{j(j+1)-n^2+\rho^2+1}
\eeq

This choice of Hilbert space is different from the approach of Alexandrov
in \cite{sergei4} who tries to build a representation for the commutation relation
at all points of space. This would correspond to an infinite refinement limit of the present
projected cylindrical function as described in the previous paragraph. The advantage
of the present approach is that the constructed functionals are truly cylindrical in that
they depend on both the connection and time normal field by a finite number of variables.

\section*{Conclusions \& Outlooks}

Projected cylindrical functions, as presented here, are a first step toward
constructing a suitable Hilbert space for covariant loop gravity.
They are properly defined in the sense that they depend on the connection
and the time normal field by a finite number of arguments. They form a Hilbert space
whose basis, the projected spin networks, can be considered as a
covariant generalisation of $SU(2)$ spin networks. We also studied
an infinite refinement limit of those spin networks in which one would know
the value of the time normal on the whole manifold instead of knowing it at
a finite number of points. It was shown that either
this limit doesn't allow an interpretation as a $L^2$ space or the refining
is not the obvious one.

\medskip

Then, it would be interesting to
fully study the introduced functionals in the framework of Alexandrov's framework
and, to start with, derive the algebra of those projected cylindrical functions
(their Dirac bracket). Moreover the use of the time normal field $\chi$ allowing
a straightforward geometrical interpretation of the states shows a clear link with
the spin foam formalism and reinforce the hope of an explicit correspondence between
the structures of loop gravity and the ones arising from the Barrett-Crane model,
even though it isn't not straightforward to derive $SU(2)$ spin networks
from the spin foam boundary states.

It would also be interesting to apply history phase space techniques
to covariant canonical loop gravity since it takes in account explicitely the embedding
(here given by the $\chi$ field) of the studied hypersurface. This could yield the searched
link with the spin foam space-time structures.

Finally, using an explicitely covariant formalism opens the door
to the systematic study of space-time related issues such as transformations
of areas under Lorentz boosts, as studied in Loop Quantum Gravity \cite{simone}.

\section*{Acknowledgments}

The author is much grateful to Daniele Oriti and Carlo Rovelli for many
stimulating discussions and comments on the manuscript.

\appendix

\section{$SO(3,1)/SO(3)$}
\label{A}

\ni
For $x$ on the future hyperboloid ${\cal H}^+$ i.e a time-like
future-oriented unit 4-vector, we can define its stabiliser
under the action $g.x$ of $SO(3,1)$:

\beq
SO(3)_x=\{h\in SO(3,1) / h.x = x\}
\eeq

\ni
The transformation law of $SO(3)_x$ under the action
of $SO(3,1)$ is

\beq
\forall g\in SO(3,1), \, SO(3)_{g.x}=gSO(3)_xg^{-1}
\label{so3tf}
\eeq

\ni
Then ${\cal H}^+$  can be seen as the quotient $SO(3,1)/SO(3)$
of $SO(3,1)$ by the {\it left} action of the subgroup $SO(3)$
- the canonical $SO(3)$ subgroup i.e the stabiliser of
$x_0=(1,0,0,0)$.
Then for $[g_0]\in  SO(3,1)/SO(3)$ and $g\in SO(3,1)$, we have:

\beq
SO(3)_{[g_0]}=\{k\in SO(3,1) / k.[g_0]=[g_0]\}
=\{k\in SO(3,1) / SO(3)g_0k= SO(3)g_0\}
\eeq

\beq
SO(3)_{g.[g_0]}=SO(3)_{[g_0g^{-1}]}=gSO(3)_{[g_0]}g^{-1}
\eeq

\ni
Now, if one considers an irreducible unitary representation $R^\I$
of $SO(3,1)$, we can decompose it on the irreducible
(finite-dimensional) representations of $SO(3)_x$
for a given vector $x$:

\beq
R^\I=\bigoplus_{j}V^j_{(x)}
\eeq

\ni
So one can choose a nice basis of $R^\I$ by considering the canonical
basis of the spaces $V^j_{(x)}$. This way, one gets a basis
$|\I x j m\rangle$. And using \Ref{so3tf}, we can related these basis
for different choices of $x$ by

\beq
|\I (g.x) j m\rangle =
D^\I(g)|\I x j m\rangle
\eeq

\section{Irreducible representations of the Lorentz group}
\label{B}

From the initial generators
of the $so(3,1)$ algebra $T_X=(A_a,-B_a)$,
we can introduce the following generators~:  
\beqs  
&H_+=iB_1-B_2, \qquad H_-=iB_1+B_2, \qquad H_3=iB_3,& \\  
&F_+=iA_1-A_2, \qquad F_-=iA_1+A_2, \qquad F_3=iA_3.&   
\eeqs 
These new generators commute in the following way:  
\beqs  
& [H_+,H_3]=-H_+, \qquad [H_-,H_3]=H_-, \qquad [H_+,H_-]=2H_3, &  
\nonumber \\  
& [H_+,F_+]=[H_-,F_-]=[H_3,F_3]=0, & \nonumber \\  
& [H_+,F_3]=-F_+, \qquad  [H_-,F_3]=F_-, & \\  
& [H_+,F_-]=-[H_-,F_+]=2F_3, & \nonumber \\  
& [F_+,H_3]=-F_+, \qquad [F_-,H_3]=F_-, & \nonumber \\  
& [F_+,F_3]=H_+, \qquad [F_-,F_3]=-H_-, \qquad [F_+,F_-]=-2H_3. &  
\nonumber       
\eeqs 
  
The irreducible infinite dimensional
representation of the Lorentz group are characterized
by two numbers $(l_0,l_1)$, where $l_0\in \Nat/2$ and $l_1\in \Cb$.   
In the space ${\cal H}_{l_0,l_1}$ of this representation one can introduce   
an orthonormal basis   
\beqs
\{ \su{j,m}\},\qquad m=-j,-j+1,\dots,j-1,j, \quad  
j=l_0,l_0+1,\dots   
\eeqs
such that the generators introduced above act in the 
following way \cite{gms}:
\beqs  
H_3\su{j,m}&=& m\su{j,m}, \nonumber \\  
H_+\su{j,m}&=& \sqrt{(j+m+1)(j-m)}\su{j,m+1}, \label{gauss-rep} \\
H_-\su{j,m}&=& \sqrt{(j+m)(j-m+1)}\su{j,m-1}, \nonumber \\  
F_3\su{j,m}&=& \gamma_{(j)}\sqrt{j^2-m^2}\su{j-1,m}+\beta_{(j)}m\su{j,m}  
-\gamma_{(j+1)}\sqrt{(j+1)^2-m^2}\su{j+1,m}, \nonumber \\  
F_+\su{j,m}&=&  
\gamma_{(j)}\sqrt{(j-m)(j-m-1)}\su{j-1,m+1}+\beta_{(j)}\sqrt{(j-m)(j+m+1)}  
\su{j,m+1} \nonumber \\  
&+& \gamma_{(j+1)}\sqrt{(j+m+1)(j+m+2)}\su{j+1,m+1}, \label{boosts-rep} \\
F_-\su{j,m}&=&  
-\gamma_{(j)}\sqrt{(j+m)(j+m-1)}\su{j-1,m-1}+\beta_{(j)}\sqrt{(j+m)(j-m+1)}  
\su{j,m-1} \nonumber \\  
&-& \gamma_{(j+1)}\sqrt{(j-m+1)(j-m+2)}\su{j+1,m-1}, \nonumber  
\eeqs 
where   
\beq 
\beta_{(j)}=-\frac{il_0l_1}{j(j+1)}, \qquad   
\gamma_{(j)}=\frac{i}{j}\sqrt{\frac{(j^2-l_0^2)(j^2-l_1^2)}{4j^2-1}}.  
\eeq 
The unitary representations correspond to two cases:   
\beqs  
1)& (n,i\rho),  \quad n\in \Nat/2, \ \rho\in \Rb & \quad - \  
principal\ series, \\  
2)& (0,\rho), \quad |\rho|<1, \ \rho\in \Rb & \quad - \ 
supplementary\ series.   
\eeqs 

The principal series are called as such for they are the ones
involved in the Plancherel decomposition of $L^2$ functions
over $SL(2,C)$ \cite{ruhl}:

\beq
f(g)=\f{1}{8 \pi^4}\sum_n \int {\rm
Tr}[F(n,\rho)D^{n,\rho}(g^{-1})](n^2+\rho^2) d\rho
\eeq
where $(n^2+\rho^2) d\rho$ is the Plancherel measure over the principal
series of representations and the Fourier transform $F$ is defined as

\beq
F(n,\rho)=\int f(g) D^{n,\rho}(g) dg
\eeq

Simple representations are the principal unitary representations
with the Casimir $-il_0l_1=n\rho=0$.
These are the representations coming into the spin foam
models as explained in \cite{bc2}.
There are obvisouly two
series of such representations: the $(n,0)$ discrete series
and $(0,\rho)$ continuous series. They both share the
fact that all their coefficients $\beta_{(j)}$ are $0$.

\section{The $SL(2,C)$ Kernel}
\label{C}

Simple representations of the $(0,\rho)$ type have a unique
$SU(2)$ invariant vector, noted $|\rho \,j=0\rangle$, and one
can define a function $K_\rho(g)$ ($K$ like {\bf K}ernel)
which is bi-invariant under $SU(2)$:

\beq
K_\rho(g)=\langle \rho \,0|
D^{(0,\rho)}(g)|\rho \,0\rangle
\eeq
Due to its invariance, it is a function of solely the boost angle
of $g$. It is useful in studying $L^2$ functions over the
upper hyperboloid ${\cal H}_+$. I merely recall basic relations
between these functions. For more details, the lector could read
\cite{bc2,finite,ruhl}.
An explicit expression of the Kernel is
\beq
K_\rho(x,y)=
K_\rho(x^{-1}y)
=\f{\sin\rho \theta}{\rho\sinh\theta}
\eeq
where $\theta$ is the hyperbolic distance (boost parameter)
between $x$ and $y$.
One can also check that
\beq
\int_0^\infty \rho^2 d\rho \, K_\rho(x,y)=
2\pi^2\delta_{{\cal H}_+}(x,y)
\eeq
where $d\mu(\rho)=\rho^2d\rho$ is the Plancherel measure
restricted to the simple representation $(n=0,\rho)$.
Moreover
\beq
\int_{{\cal H}_+}dy K_\rho(x,y)
K_{\rho'}(y,z)=\f{\delta(\rho-\rho')}{\rho^2}
K_\rho(x,z)
\eeq
which means that the Kernel $K$ is a projection
onto the $\rho$ Fourier component on the hyperboloid
${\cal H}_+$.


\begin{thebibliography}{99}

\bibitem{spinnet}
Laurent Freidel, Etera R Livine,
{\it Spin Networks for Non-Compact Groups},
hep-th/0205268

\bibitem{al}
Abhay Ashtekar, Jerzy Lewandowski,
{\it Projective Techniques and Functional Integration},
J.Math.Phys. 36 (1995) 2170-2191,
gr-qc/9411046

\bibitem{carlo}
Carlo Rovelli,
{\it Ashtekar for,ulation of general relativity
and loop space nonperturbative quantum gravity: A report},
Class. Quant. Grav. 8 (1991) 1613-1676


\bibitem{simple}
Laurent Freidel, Kirill Krasnov,
{\it Simple Spin Networks as Feynman Graphs},
J.Math.Phys. 41 (2000) 1681-1690,
hep-th/9903192

\bibitem{freidel}
Roberto De Pietri, Laurent Freidel,
{\it $so(4)$ Plebanski Action and Relativistic Spin Foam Model},
Class.Quant.Grav. 16 (1999) 2187-2196,
gr-qc/9804071


\bibitem{bc1}
John W Barrett, Louis Crane,
{\it Relativistic spin networks and quantum gravity},
J.Math.Phys. 39 (1998) 3296-3302,
gr-qc/9709028 


\bibitem{bb}
John C Baez, John W Barrett,
{\it The Quantum Tetrahedron in 3 and 4 Dimensions},
Adv.Theor.Math.Phys. 3 (1999) 815-850,
gr-qc/9903060




\bibitem{bc2}
John W Barrett, Louis Crane,
{\it A Lorentzian Signature Model for Quantum General Relativity},
Class.Quant.Grav. 17 (2000) 3101-3118,
gr-qc/9904025

\bibitem{alejl1}
Alejandro Perez, Carlo Rovelli,
{\it Spin foam model for Lorentzian General Relativity},
Phys.Rev. D63 (2001) 041501,
gr-qc/0009021

\bibitem{alejl2}
Alejandro Perez, Carlo Rovelli,
{\it 3+1 spinfoam model of quantum gravity with
spacelike and timelike components},
Phys.Rev. D64 (2001) 064002,
gr-qc/0011037

\bibitem{finite}
John C Baez, John W Barrett,
{\it Integrability for Relativistic Spin Networks},
Class.Quant.Grav. 18 (2001) 4683-4700,
gr-qc/0101107



\bibitem{etera1}
Etera R Livine,
{\it Immirzi parameter in the Barrett-Crane model?},
gr-qc/0103081

\bibitem{etera2}
Etera R Livine, Daniele Oriti,
{\it Barrett-Crane spin foam model from generalized BF-type action
for gravity},
Phys.Rev. D65 (2002) 044025,
gr-qc/0104043

\bibitem{etera3}
Etera R Livine, Daniele Oriti,
{\it Spin Foam Space-Time as Quantum Causal Set},
in preparation



\bibitem{daniele}
Daniele Oriti,
{\it Spacetime geometry from algebra: spin foam models
for non-perturbative quantum gravity},
Rept.Prog.Phys. 64 (2001) 1489-1544,
gr-qc/0106091

\bibitem{sergei1}
Sergei Alexandrov,
{\it SO(4,C)-covariant Ashtekar-Barbero gravity and the Immirzi
parameter},
Class.Quant.Grav. 17 (2000) 4255-4268,
gr-qc/0005085

\bibitem{sergei2}
Serguei Alexandrov, Dmitri Vassilevich,
{\it Area spectrum in Lorentz covariant loop gravity},
Phys.Rev. D64 (2001) 044023,
gr-qc/0103105

\bibitem{sergei3}
Sergei Alexandrov,
{\it On choice of connection in loop quantum gravity},
Phys.Rev. D65 (2002) 024011,
gr-qc/0107071

\bibitem{sergei4}
Serguei Alexandrov,
{\it Hilbert Space for Covariant Loop Quantum Gravity},
gr-qc/0201087

\bibitem{su2fromcov}
Sergei Alexandrov, Etera R Livine,
{\it $SU(2)$ Loop Quantum Gravity from Covariant Theory},
in preparation


\bibitem{holst}
S{\"o}ren Holst,
{\it Barbero's Hamiltonian derived from a generalized Hilbert-Palatini action},
Phys.Rev. D53 (1996) 5966-5969,
gr-qc/9511026

\bibitem{barros}
 Nuno Barros e Sa,
{\it Hamiltonian analysis of General relativity with the Immirzi parameter},
Int.J.Mod.Phys. D10 (2001) 261-272,
gr-qc/0006013


\bibitem{simone}
 Carlo Rovelli, Simone Speziale,
{\it Reconcile Planck-scale discreteness and the
Lorentz-Fitzgerald contraction},
gr-qc/0205108

\bibitem{thiemann}
 Thomas Thiemann,
{\it  Introduction to Modern Canonical Quantum General Relativity},
gr-qc/0110034 


\bibitem{knapp}
A. W. Knapp
{\it Representation of semi-simple groups},
Princeton University press.

\bibitem{ruhl}
W Ruhl, ``The Lorentz Group and Harmonic Analysis" (WA Benjamin
Inc, New York 1970)

\bibitem{gms}
I.~M.~Gel'fand, R.~A.~Minlos and Z.~Ya.~Shapiro, {\it Representations
of the rotation and Lorentzi groups and their applications}
(Pergamon Press, 1963), pages 187-189

\end{thebibliography}
\end{document}